\documentstyle[preprint,aps,psfig]{revtex}
\tightenlines

\begin{document}
\draft \preprint{MKPH-T-97-20} 
\title{On the extraction of electromagnetic properties of the 
$\Delta$(1232) excitation from pion photoproduction}

\author{Th.\ Wilbois}
\address
{Institut f\"ur Theoretische Physik, Universit\"at Hannover, D-30167 Hannover,
Germany}
\author{P.\ Wilhelm and H.\ Arenh\"ovel}
\address
{Institut f\"ur Kernphysik, Johannes Gutenberg-Universit\"at, 
D-55099 Mainz,Germany}
\maketitle

\begin{abstract}
Several methods for the treatment of pion photoproduction in the region of
the $\Delta(1232)$ resonance are discussed, in particular the effective
Lagrangian approach and the speed plot analysis are compared to a
dynamical treatment. As a main topic, we discuss the extraction of the
genuine resonance parts of the magnetic dipole and electric quadrupole
multipoles of the electromagnetic excitation of the resonance. To this
end, we try to relate the various values for the ratio $R_{EM}$ of the E2
to M1 multipole excitation strengths for the $\Delta(1232)$ resonance as
extracted by the different methods to corresponding ratios of a dynamical
model. Moreover, it is confirmed that all methods for extracting resonance
properties suffer from an unitary ambiguity which is due to some
phenomenological contributions entering the models. 
\end{abstract}


\section{Introduction}\label{sect1}

One of the most prominent manifestations of internal nucleon structure
is the $\Delta$(1232) resonance which is seen, e.g., in pion
scattering and electromagnetic (e.m.) pion production. Thus it is not
surprising that a large number of experimental and theoretical
investigations has been devoted to the understanding of the structure
of this resonance. Recently, the question of the relative strength of
the electric quadrupole (E2) to the magnetic dipole (M1) excitation in
the e.m.\ $\gamma N\leftrightarrow \Delta$ transition has become one
of the central questions and is subject of numerous papers
\cite{TaO85,Yan85,DaM91,BeN93,WiW96,BuH97,LuT97,ShD97}.  While the M1
strength in the simplest approach is directly related to the magnetic
properties of the constituents, the E2 strength may be interpreted as
a measure of internal spatial deformation, i.e., as deviation of the
orbital wave functions from spherical symmetry. Thus, there is
considerable experimental effort to provide accurate data on pion
photo- and electroproduction from the nucleon in the $\Delta$
resonance region \cite{Bla92,BeK97}. Various methods have been
proposed in the literature for the isolation of the genuine resonant
part of the experimentally measured multipoles which, however, give
quite different answers \cite{TaO85,DaM91,NoB90,HaD96,SaL96}.

The general problem of such an extraction lies in the fact 
that the experimentally observable multipoles contain important
background contributions which cannot be separated in a simple and unique 
way, so that in most explicit evaluations a specific model is used. Thus it 
is not surprising that the different prescriptions result in different 
values of resonance properties. In fact, for photoproduction the values for 
$R_{EM}$, the ratio of electric quadrupole to magnetic dipole, 
vary between $+4\,\%$ to $-8\,\%$ (see e.g.\ the discussion in \cite{BeN93}). 
However, in view of 
the various methods used, it remains unclear how these values 
should be compared to each other. In other words, is there a common basis 
for a meaningful comparison? 

It is the purpose of this paper to address this question in greater detail. 
To this end we will compare three different methods which 
have been recently applied.  They are based on (i) a dynamical model (DM) 
for the coupled $\pi$N-$\Delta$-$\gamma$N system, (ii) the effective
Lagrangian approach (EL), and (iii) the recently proposed 
speed plot analysis (SP). We will take the DM as a common point of 
reference for this study. 

To this end we briefly review in Sect.\ \ref{sect2} the treatment of
pion photoproduction based on a dynamical model.  Special attention is
paid to the general structure of the pion nucleon scattering and
photoproduction amplitudes and the implication of unitarity. Sect.\
\ref{sect3} collects the basic formulas of the treatment of these
reactions in the framework of ELs, and the connection between
quantitities like resonance mass, coupling constants of the ELs and
the corresponding quantities of a DM is derived. Moreover, a numerical
comparison is performed.  In Sect.\ \ref{sect4} we briefly discuss the
SP and point out some shortcomings in the present application and how
they can be avoided.  The paper ends with a summary and some
conclusions.

\section{Dynamical model}
\label{sect2}

There are two main advantages of treating e.m.\ pion production 
dynamically, i.e., solving for a given $\pi N$ interaction the 
corresponding Lippman-Schwinger equation. First of all, it ensures that all 
two-body 
unitarity constraints are automatically respected. Secondly, it provides a 
well defined basis for the analysis of the role of the final state interaction 
in the e.m.\ reaction. In such a dynamical approach the model parameters 
are usually adjusted in order to reproduce all available on-shell data. At 
present the main disadvantage lies in the fact that a dynamical model requires 
still some purely phenomenological input, e.g., form factors in 
order to regularize the driving terms 
of the interaction, which is essential to solve the dynamical equations. 
Further problems are related to the neglect of three-body unitarity above 
two-pion threshold and to the requirements of relativity and gauge 
invariance. However, various improvements have been achieved recently 
\cite{NoB90,SaL96,GrS93,HuY94,PaT97}. 
Notwithstanding these shortcomings, such a dynamical model still provides 
the most satisfactory and comprehensive description. 

We shall first briefly review the salient features of a dynamical
model for the coupled $\pi$N-$\Delta$-$\gamma$N system following the
notation of \cite{TaO85,WiW96}. The Hilbert space is assumed to be a
direct sum of bare $\Delta$, $\pi N$ and $\gamma N$ states with
corresponding projectors $P_\Delta$, $P_{\pi N}$, and $P_{\gamma
N}$. The Hamiltonian is taken as
\begin{equation}
h = h_0(m_\Delta^0 ) + v_{\pi\pi}^B + \Big( v_{\Delta \pi N} 
+ v_{\pi\gamma N}^B +v_{\Delta \gamma N} 
+h.c.\Big) \, ,
\label{eqn:h}
\end{equation}
with a background $\pi N$ interaction $v_{\pi\pi}^B=P_{\pi N} (h-h_0)
P_{\pi N}$, the $\pi N\Delta$ vertex $v_{\Delta\pi N}=P_\Delta h
P_{\pi N}$, a nonresonant $\gamma N\rightarrow \pi N$ driving term
$v_{\pi\gamma N}^B=P_{\pi N} h P_{\gamma N}$, and the $\gamma N\Delta$
vertex $v_{\Delta \gamma N} =P_\Delta h P_{\gamma N}$.  One should
note that the energy $h_0$ in the $\Delta$ sector depends on the bare
resonance mass $m_\Delta^0$ which is a model parameter to be fitted.

\begin{figure}
\centerline{\psfig{file=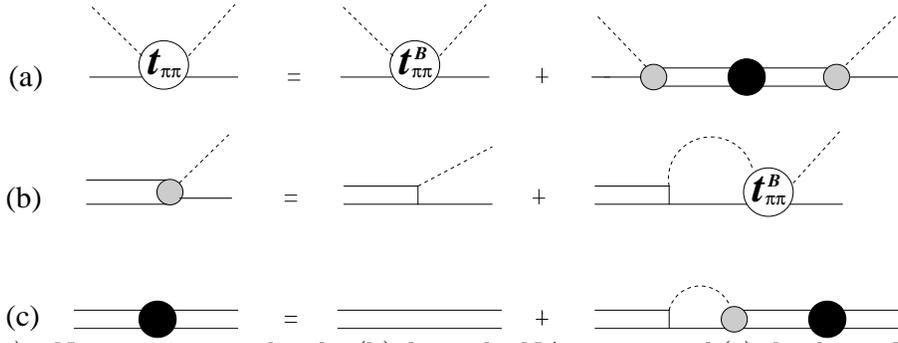,width=12cm,angle=0}}
\caption{(a) $\pi N$ scattering amplitude, (b) dressed $\pi N\Delta$ vertex,
and (c) the dressed $\Delta$ propagator $g_\Delta$. }
\label{fig:pipi}
\end{figure}

Within such a model the elastic $\pi N$ scattering amplitude
$t_{\pi\pi}$ can be split into a background term $t_{\pi\pi}^{B}$ 
and a resonant part $t_{\pi\pi}^{R}$ 
\begin{equation}
t_{\pi\pi} = t_{\pi\pi}^{B} + t_{\pi\pi}^{R}\,.
\label{eqn:tpipi}
\end{equation}
The background term is defined by the integral equation 
\begin{equation}
t_{\pi\pi}^{B} = v_{\pi\pi}^{B} +
v_{\pi\pi}^{B}\, g_{\pi N}^0 \,t_{\pi\pi}^{B} 
\end{equation}
with the bare $\pi N$ propagator $g_{\pi N}^0$. 
The resonant part 
\begin{equation}
t_{\pi\pi}^{R} = 
\tilde v _{\pi N \Delta} g_\Delta \tilde v _{\Delta \pi N} 
\label{eqn:tpipipr}
\end{equation}
is determined by the dressed $\pi N\Delta$ and $\Delta \pi N$ vertices 
\begin{equation}
\tilde v_{\Delta \pi N} = v_{\Delta \pi N} + v_{\Delta \pi N} 
g_{\pi N}^0 t_{\pi\pi}^{B}, \qquad
\tilde v_{\pi N \Delta}= v _{\pi N \Delta} + t_{\pi\pi}^{B} 
g_{\pi N}^0 v _{\pi N \Delta} ,
\label{eqn:vpitilde}
\end{equation}
and the dressed $\Delta$ propagator $g_\Delta$, obtained from 
\begin{equation}
g_\Delta (s) = g_{\Delta}^0(s)  + g_{\Delta}^0(s)  \Sigma_\Delta(s)  
g_\Delta(s)  ,
\label{eqn:gdeltadress}
\end{equation}
with the free $\Delta$ propagator $g_\Delta^0(z)=(z-m_\Delta^0)^{-1}$ and 
the $\Delta$ self energy 
\begin{equation}
\Sigma _\Delta (s) = \tilde v_{\Delta \pi N}
g_{\pi N}^0 v_{\pi N \Delta} = v_{\Delta \pi N} g_{\pi N}^0
\tilde v_{\pi N \Delta}\,.
\label{eqn:Sigma}
\end{equation}
The equations (\ref{eqn:tpipi}) and
(\ref{eqn:tpipipr})-(\ref{eqn:gdeltadress}) are illustrated
diagrammatically in Fig.\ \ref{fig:pipi}.  We have already mentioned
the fact that the driving terms of the background and resonance
amplitudes contain at present phenomenological ingredients which are
fitted to the experimental data. Thus these amplitudes and the
corresponding phases are to some extent model dependent. Recently we
have shown that this model dependence can be interpreted as different,
unitarily equivalent representations, yielding different splittings
into background and resonant amplitudes but leaving the total
amplitude unchanged \cite{WiW96}.

\begin{figure}
\centerline{\psfig{file=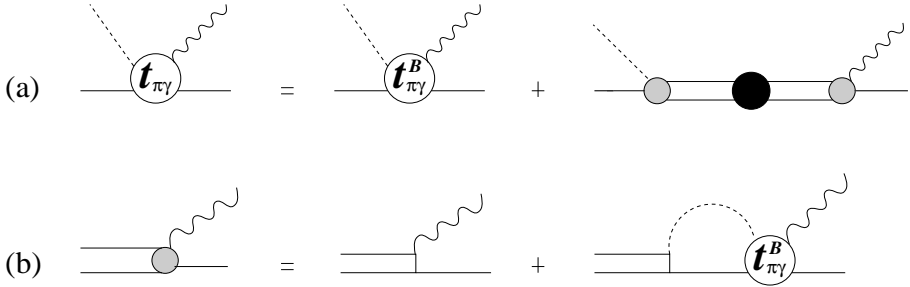,width=12cm,angle=0}}
\caption{(a) Pion production  amplitude and (b) dressed $\gamma N\Delta$  
vertex.}
\label{fig:pigamma}
\end{figure}

The structure of the photoproduction amplitude 
\begin{equation}
t_{\pi\gamma} = t_{\pi\gamma}^{B} + t_{\pi\gamma}^{R}
\label{eqn:tpigamma}
\end{equation}
is completely analogous.
It again splits into a background $t_{\pi\gamma}^B$, determined by 
\begin{equation}
t_{\pi\gamma}^B =  
v_{\pi\gamma N}^B +t_{\pi\pi}^B g_{\pi N}^0 v_{\pi\gamma N}^B\,, 
\end{equation}
and a resonant part 
\begin{equation}
t_{\pi\gamma}^R = \tilde v_{\pi N \Delta} g_\Delta 
\tilde v_{\Delta \gamma N}\,,
\end{equation}
referred to as ``dressed'' resonant amplitude. It 
contains the dressed $\gamma N\Delta$ vertex
\begin{equation}
\tilde v_{\Delta \gamma N} = v_{\Delta \gamma N} + 
v_{\Delta \pi N} g_{\pi N}^0 t_{\pi\gamma}^{B}\,.
\label{eqn:vgammatilde}
\end{equation}
Accordingly,  one may split $t_{\pi\gamma}^R $ further into the so-called
bare resonant part $t_{\pi\gamma}^\Delta$ and a vertex renormalization 
$t_{\pi\gamma}^{VR}$, i.e., 
\begin{eqnarray}
t_{\pi\gamma}^R &=& t_{\pi\gamma}^\Delta + t_{\pi\gamma}^{VR}\,,
\end{eqnarray}
where
\begin{eqnarray}
t_{\pi\gamma}^\Delta & = & \tilde v_{\pi N\Delta} g_\Delta 
v_{\Delta\gamma N}\,,
\label{tpigdelta}
\end{eqnarray}
and
\begin{eqnarray}
t_{\pi\gamma}^{VR} & = & \tilde v_{\pi N\Delta} g_\Delta v_{\Delta\pi N} 
g_{\pi N}^0 t_{\pi\gamma}^B\,.
\end{eqnarray}
The diagrammatic representation of $t_{\pi\gamma}$ is shown in 
Fig.\ \ref{fig:pigamma}. It is obvious that for both multipole 
amplitudes the same formal structure holds. 

The form of the amplitude as discussed 
above is quite general
below the $\pi\pi$ production threshold. In that energy regime, the
coupling to the closed $\pi\pi N$ channel may be effectively incorporated 
into a {\em real}, energy dependent background interaction, 
where the energy dependence arises through the
energy dependence of the propagators in the intermediate $\pi\pi N$
states.  Note that such an energy dependence is already introduced by
considering the crossed nucleon pole term as background mechanism
$v_{\pi\pi}$ in the (3,3) channel.

We turn now to the discussion of the basic relations between on-shell
matrix elements and phase shifts. In order to avoid a new notation, we will 
understand all amplitudes in connection with the phase shifts as 
the corresponding on-shell quantities. For $\pi N$
scattering, the scattering amplitude reads
\begin{equation}
t_{\pi\pi} = -\frac{1}{\rho} e^{i\delta} \sin \delta\,,
\end{equation}
with a phase space factor $\rho$ determined by the adopted
normalization conventions, i.e., the scattering matrix is given by
\begin{equation}
s_{\pi\pi} = 1 + 2 i \rho t_{\pi\pi}\,.
\end{equation}
The usual choice is $\rho = - q$, where $q$ denotes the on-shell
momentum of the pion in the $\pi N$ c.m.\ frame.
Since the background is unitary by itself, one can define a
nonresonant background phase shift $\delta_B^{DM}$ by
\begin{equation}
t^B_{\pi\pi} = -\frac{1}{\rho} e^{i\delta _B^{DM}} \sin \delta_B^{DM}\,.
\end{equation}
One then easily deduces for the resonant part
\begin{equation}
t_{\pi\pi}^R = -\frac{1}{\rho} e^{i(\delta + \delta_B^{DM})} 
\sin \delta_R^{DM}\,,
\label{eqn:tpipiron}
\end{equation}
where the resonance phase shift $\delta_R^{DM}=\delta-\delta_B^{DM}$ 
is given by 
\begin{equation}
\tan{\delta_R^{DM}}(s)=\frac{\Im m \Sigma_\Delta(s)}
{\sqrt{s}- m_\Delta(s) }\,.\label{tandr}
\end{equation}
Here we have introduced the dressed $\Delta$ mass
\begin{equation}
m_\Delta(s) = m_\Delta^0 + \Re e \Sigma_\Delta(s) \,.\label{dressedmass}
\end{equation}

Neglecting small Compton scattering corrections \cite{BeM92}, Watson's 
theorem states that the phase of the photoproduction amplitude below the 
two pion threshold is given by the hadronic $\pi N$ scattering phase shift 
$\delta$, i.e., 
\begin{equation}
t_{\pi\gamma} = \bar t_{\pi\gamma}  e^{i\delta}
\end{equation}
where $\bar t_{\pi\gamma}$ is real. Similarly, one has for the background 
contribution 
\begin{equation}
t_{\pi\gamma}^B = \bar t_{\pi\gamma}^B e^{i\delta_B^{DM}}\,,
\end{equation}
with the $\pi N$ scattering background phase shift $\delta_B^{DM}$ and a real 
$\bar t_{\pi\gamma}^B$. 
It is convenient to define an additional phase $\delta_p^{DM}$, called 
photoproduction background phase \cite{TaO85}, so that 
\begin{equation}
t_{\pi\gamma}^R  =  \bar t_{\pi\gamma}^R e^{i(\delta+ \delta_p^{DM})}\,,
\label{eqn:deltapdef}
\end{equation}
where again $\bar t_{\pi\gamma}^R$ is real. One should keep in mind that 
this phase $\delta_p^{DM}$ is in general different for the different 
multipole amplitudes. Moreover, $\delta_p^{DM}$ is determined by the elementary
interaction model.  
In detail, one has 
\begin{equation}
\tan{\delta_p^{DM}}=\frac{\sin{\delta_R^{DM}}}
{\bar t_{\pi\gamma}-\bar t_{\pi\gamma}^B \cos{\delta_R^{DM}}}\,.
\end{equation}
For the comparison with the ELs it is useful to express the modulus of the 
total photoproduction amplitude in terms of the modulus of the resonance 
amplitude and the photoproduction background phase 
\begin{equation}
\label{eqn:tpig}
\bar t_{\pi\gamma}=\bar t_{\pi\gamma}^R
\frac{\sin{(\delta_R^{DM}+\delta_p^{DM})}}{\sin{\delta_R^{DM}}}\,.
\end{equation}
Furthermore, for the moduli of the background and bare resonance 
amplitudes, one easily finds the following relations
\begin{eqnarray}
\label{eqn:tpigb}
  \bar t_{\pi\gamma}^B & = & \bar t_{\pi\gamma}^R 
            \frac{\sin\delta_p^{DM}}{\sin \delta_R^{DM} }\,,\\
\label{eqn:tpigd}
  \bar t_{\pi\gamma}^\Delta &=&  -\frac{v_{\Delta\gamma N}}
{\rho \bar {\tilde v}_{\Delta \pi N}} \sin\delta_R^{DM}\,.
\end{eqnarray}
Note that the bare amplitude $t_{\pi\gamma}^\Delta$ carries the full phase
$\delta$ (see (\ref{tpigdelta})). An expression, alternative to 
(\ref{eqn:tpig}), is then
\begin{equation}
\label{eqn:tpigalt}
\bar t_{\pi\gamma}=\bar t_{\pi\gamma}^R \cos \delta_p^{DM}
                   +\bar t_{\pi\gamma}^B \cos \delta_R^{DM}\,,
\end{equation}
which exhibits the resonance and background contributions to 
$\bar t_{\pi\gamma}$. 

At the end of this brief review, we collect different possible definitions 
of $E2/M1$ ratios $R_{EM}$ for the $\gamma N \Delta$ transition. 
The ratio of the full amplitudes
\begin{equation}
R_{EM} = \frac{t_{\pi\gamma}(E2)}{t_{\pi\gamma}(M1)} 
= \frac{\bar t_{\pi\gamma}(E2)}{\bar t_{\pi\gamma}(M1)}
\end{equation}
is directly related to the experimentally observable multipoles 
which, according to Watson's theorem, is a real but energy
dependent quantity.  The bare ratio 
\begin{equation}
R^\Delta_{EM}(\mbox{DM}) = 
  \frac{t_{\pi\gamma}^\Delta(E2)}{t_{\pi\gamma}^\Delta(M1)}
=  \frac{\bar t_{\pi\gamma}^\Delta(E2)}{\bar t_{\pi\gamma}^\Delta(M1)}
\end{equation}
is also a real number because the bare amplitudes carry the total phase 
as mentioned above. It is directly related to the ratio of the strengths of
the coupling constants in the $\gamma N\Delta$ transition current 
and thus is energy dependent only if the coupling is energy dependent. 

With respect to the ratio of the dressed multipoles, one may define 
two different quantities, since the dressed multipoles carry different 
phases, namely the complex quantity 
\begin{equation}
 R_{EM}^R (\mbox{DM}) = \frac{t_{\pi\gamma}^R(E2)}{t_{\pi\gamma}^R(M1)}\,,
\end{equation}
or, alternatively, the ratio of the moduli 
\begin{equation}
\tilde R_{EM}^R (\mbox{DM}) = 
  \frac{\bar t_{\pi\gamma}^R(E2)}{\bar t_{\pi\gamma}^R(M1)}
\,.\label{REMtilde}
\end{equation}
As we will see in the next Section, this latter quantity is closely 
related to $R_{EM}$ defined in the EL. Note, that both dressed ratios
are energy dependent. 

\section{Effective Lagrangian approach and unitarization methods}
\label{sect3}

Let us first summarize the treatment of pion scattering and pion 
photoproduction in the
effective Lagrangian approach (EL). Here, only the lowest order tree level 
Feynman diagrams are considered including background and resonance 
contributions. Subsequently, the resulting amplitude is 
unitarized. Three different prescriptions for unitarization 
have been proposed in the literature, the Olsson method
\cite{Ols74,Ols77b}, the Noelle method \cite{Noe74} and the K-matrix approach 
\cite{DaM91}. 
The dependence of the photoproduction process on the adopted
unitarization scheme has already been extensively studied by Davidson,
Mukhopadhyay and Wittman \cite{DaM91}, and we follow their notation for 
convenience. 

We start by collecting the basic formulas for pion nucleon scattering in the 
energy region of the $\Delta$ resonance. 
The different unitarization methods can be parametrized by the
following relation for the phase shift 
\begin{equation}
\tan \delta = \frac{1+\epsilon \tan \delta_B^{EL}}{\epsilon + \eta
\tan\delta_B^{EL}} \,,\label{tand}
\end{equation}
where $\tan{\delta_B^{EL}}$ is given by the nonresonant tree diagrams. 
The quantity $\epsilon$ is interpreted as the $s$-channel
contribution of the $\Delta$ resonance and parametrized in the form
\begin{equation}
\epsilon (s)= \frac{M_\Delta^2-s}{M_\Delta\Gamma_\Delta(s)}\,,
\label{eqn:epsdef}
\end{equation}
where the constant mass $M_\Delta$ is obtained as the zero of $\epsilon(s)$. 
The parameter $\eta$ 
in (\ref{tand}) distinguishes the different unitarization methods,
i.e., $\eta =-1,0,1$ refers to the Noelle, K-matrix and Olsson approach,
respectively. Obviously, all methods coincide
in the case of a vanishing background phase shift $\delta_B^{EL}=0$. 
For the decay width, the following ansatz is used 
\begin{equation}
\Gamma_\Delta (s) =\frac{g_{\pi N\Delta}^2(s) q^3(s)(E_f(s)+M)
                   (\sqrt{s}+M_\Delta)}{24\pi m_\pi^2 \sqrt{s}M_\Delta},
\label{eqn:gamdef}
\end{equation}
where $q(s)$ and $E_f(s)$ denote respectively the pion momentum and the 
nucleon energy in the $\Delta$ rest frame. Furthermore, 
a weak energy dependence of the coupling constant 
$g_{\pi N\Delta}$ is allowed for 
\begin{equation}
  g_{\pi N\Delta}(s) = g_{\pi N\Delta}(M_\Delta^2) +
  \frac{B(s-M_\Delta^2)}{M_\Delta^2} + \frac{C(s-M_\Delta^2)^2}{M_\Delta^4},
\label{eqn:gpinddef}
\end{equation}
with free parameters $B$ and $C$. The background phase shift is 
parametrized as
\begin{equation}
\tan\delta_B^{EL} = a \left(\frac{q}{m_\pi} \right) ^3
             + b \left(\frac{q}{m_\pi} \right) ^5,
\end{equation}
with free parameters $a$ and $b$ to be fitted as well. 

In order to establish the connection between the EL and the DM one just has to
cast $\tan{\delta}$ of the DM into a form which corresponds to 
(\ref{tand}) in terms of the corresponding dynamical quantitities. 
A straightforward calculation gives 
\begin{equation}
\tan \delta = \frac{1+\mu \tan \delta_B^{DM} }{\mu - \tan \delta_B^{DM} }\,,
\label{tandeldm}
\end{equation}
with
\begin{equation}
\mu = \frac{1}{\tan{\delta_R^{DM}}}\,,
\end{equation}
where $\tan\delta_R^{DM}$ is given in (\ref{tandr}). 

Equating now (\ref{tandeldm}) with (\ref{tand}), one can express $\epsilon 
(s )$ as function of the DM variables $\delta_R^{DM}$ and $\delta_B^{DM}$, 
the EL background phase $\delta_B^{EL}$ and the unitarization parameter 
$\eta$
\begin{eqnarray}
\epsilon (s ) & = & \frac {\mu (1 -\eta \tan \delta_B^{DM} 
\tan \delta_B^{EL} ) - (\tan \delta_B^{DM} +\eta \tan \delta_B^{EL} )}
{1+ \mu(\tan \delta_B^{DM}-\tan \delta_B^{EL} )+ 
\tan \delta_B^{DM}\tan \delta_B^{EL}}
\,.\label{eqn:epseta}
\end{eqnarray}

By means of Eq.\ (\ref{eqn:epsdef}), the EL quantitities $M_\Delta$ and 
$\Gamma_\Delta$ can be expressed through DM quantities. In addition, they 
will also depend on the EL background phase $\delta_B^{EL}$ and the 
unitarization parameter $\eta$. For example, one finds for the resonance 
mass parameter
\begin{equation}
M_\Delta=m_\Delta + \frac{\tan \delta_B^{DM} +\eta 
\tan \delta_B^{EL}}{1 -\eta \tan \delta_B^{DM}\tan \delta_B^{EL}}
\,\Im m\,\Sigma_\Delta \,,
\end{equation}
where the right hand side is to be taken at $s_0=M_\Delta^2$ 
and the dressed energy dependent mass $m_\Delta$ is defined in 
(\ref{dressedmass}). 
The dependence on both the EL background phase $\delta_B^{EL}$ and the DM 
background phase $\delta_B^{DM}$ reflects the representation dependence 
mentioned above in the discussion of the DM. 

If now we assume the same representation, i.e., identify the background phases 
$\delta_B=\delta_B^{EL}=\delta_B^{DM}$, then the expressions simplify 
considerably, yielding
\begin{eqnarray}
\epsilon (s ) & = & \frac {\mu (1 -\eta \tan ^2 \delta_B ) 
-\tan \delta_B  (1 +\eta )}{1+\tan^2\delta_B }\,,\label{epsilons}\\
M_\Delta &=& m_\Delta + \frac{(1 +\eta)\tan \delta_B}
{1 -\eta \tan^2 \delta_B}\,\Im m\,\Sigma_\Delta \,.\label{mdeltas}
\end{eqnarray}
This is an implicit equation for the mass $M_\Delta$ in the 
respective unitarization scheme. One has in detail
\begin{eqnarray}
M_\Delta & = & m_\Delta (M_\Delta) \hspace{3cm} \mbox{(Noelle)}\,,
\label{eqn:M_dn}\\
M_\Delta & = & m_\Delta (M_\Delta) + \Im m \Sigma_\Delta (M_\Delta)\qquad 
\mbox{(K-matrix)}\label{eqn:M_dk}\,,\\
M_\Delta & = & m_\Delta (M_\Delta) + 2 \frac{\Im m \Sigma_\Delta (M_\Delta)}
                                 {1-\tan^2\delta_B(M_\Delta)}\qquad
                              \mbox{(Olsson)}\label{eqn:M_do}\\
         & = & m_\Delta (M_\Delta) + 2\,\Im m \Sigma_\Delta (M_\Delta) 
               +{\cal O}(\tan^2 \delta_B)\nonumber \,.
\end{eqnarray}
It is easily seen that in the case of the K-matrix method, $M_\Delta$
corresponds to the energy at which the full
phase shift $\delta$ is equal to $\pi/2$. Note that the sign of $\Im m
\Sigma_\Delta$ is not fixed a priori. For the model B of Tanabe and
Ohta, upon which our later comparison is based, one obtains a negative
sign and thus within this model $M_\Delta$(Olsson) $< M_\Delta$(K-matrix) $<
M_\Delta$(Noelle).
 
Turning now to pion photoproduction, we begin with a brief 
summary of the treatment of this process as given in \cite{DaM91}.
In the EL, one starts from the tree level contributions of background and 
resonance terms
\begin{equation}
t_{\pi\gamma}(\mbox{tree}) = A_B + \frac{N}{\epsilon}.\label{eltree}
\end{equation}
For our purpose, it is not necessary to specify the explicit expressions for
the real numbers $A_B$ and $N$, which are given in  \cite{DaM91}. The 
unitarized amplitude then has the form
\begin{equation}
t_{\pi\gamma}=\bar t_{\pi\gamma}e^{i\delta}\,,
\end{equation}
with
\begin{equation}
\bar t_{\pi\gamma}= \tilde N \sin{(\delta_R^{EL}+\delta_p^{EL})}\,.
\label{tpigel}
\end{equation}
The real amplitude $\tilde N$ and the photoproduction background phase 
$\delta_p^{EL}$ are functions of $N$, $A_B$ and the phases $\delta_R^{EL}$ 
and $\delta_B^{EL}$ whose explicit form depends on the unitarization method 
and which will be given below. The ratio $R^{R}_{EM}$ is then given in terms 
of the appropriate electric and magnetic amplitudes $N_{E/M}$ defined 
correspondingly as in (\ref{eltree}) 
\begin{equation}
R^{R}_{EM}(\mbox{EL}):= \frac{N_E(\mbox{EL})}{N_M(\mbox{EL})}\,.
\end{equation}
In the literature, the ratios $R^R_{EM}$(EL) are quoted just at
the energy $M_\Delta$ of the respective unitarization scheme.

Now we can establish the relations between the EL amplitudes $N_{E/M}$ and 
the DM quantities on the basis that the total $\pi N$ phase shift $\delta$ and 
the total multipole amplitudes can be identified, i.e., give a satisfactory 
fit to experimental data. 
In Olsson's method, the unitarized amplitude is given in the form 
\begin{equation}
\bar t_{\pi\gamma}=A_B\cos \delta_R^{EL}+ N\sin \delta_R^{EL}\cos \delta_p^{EL}
\,,
\end{equation}
with
\begin{equation}
\sin \delta_p^{EL} = \frac{A_B}{N}\,,
\label{eqn:dmwdeltap}
\end{equation}
leading to the correspondence $\tilde N (\mbox{Olsson})=N$.  
As already pointed out by Olsson \cite{Ols74}, no solution arises in
the case when $|A_B| > |N|$, as is the case for the $E2$ excitation of
the $\Delta$.  This problem is avoided, when the helicity amplitudes
are unitarized instead of the multipoles \cite{DaM91}. However, this
trick would not work in the case of a weak resonance which is produced
by one e.m.\ multipole only.  The comparison with the DM,
which has already been performed by Tanabe and Ohta \cite{TaO85}, is
based on the identification of (\ref{tpigel}) with (\ref{eqn:tpig}).
It yields the following relation
\begin{equation}
N= \frac{\sin{(\delta_R^{DM}+\delta_p^{DM})}}{\sin{(\delta_R^{EL}+\delta_p^{EL
})}}\,\frac{\bar t_{\pi\gamma}^R}{\sin \delta_R^{DM}}\,.
\end{equation}
It is worth noting that, when identifying the respective background
contributions in EL and DM, the case $|A_B| > |N|$ {\em cannot} appear, 
because in the dynamical model the photoproduction background 
phase $\delta_p^{DM}$ is a well defined quantity.
Although this relation between $N$ and the dressed amplitude 
$\bar t_{\pi\gamma}^R$ depends also on the various phases of DM and EL and 
thus does not allow a simple interpretation, 
one sees that the ratio $R_{EM}$ is just 
the ratio of the moduli of the dressed multipoles in the DM 
(see (\ref{REMtilde}))
\begin{equation}
R^R_{EM}(\mbox{Olsson}) = 
 \frac{\bar t_{\pi\gamma}^R (E2)}{\bar t_{\pi\gamma}^R (M1)}=
  \tilde R^R_{EM}(\mbox{DM})\,.
\end{equation}

In Noelle's scheme, the unitarized amplitude is given by 
\begin{equation}
\bar t_{\pi\gamma}=A_B\cos \delta + N\sin \delta_R^{EL}\,,
\label{tpignoelle}
\end{equation}
which leads to the following expressions 
\begin{eqnarray}
\tilde N^2&=&N^2\cos^2\delta_B^{EL}+(N\sin \delta_B^{EL}-A_B)^2\,,\\
\tan \delta_p^{EL}&=&\frac{A_B \cos \delta_B^{EL}}
{N -A_B\sin \delta_B^{EL}}\,.
\end{eqnarray}
The connection between $N$ and the DM amplitudes reads
\begin{equation}
N(\mbox{Noelle}) = \frac{\bar t_{\pi\gamma}\cos\delta_B^{EL}- 
\bar t_{\pi\gamma}^B \cos \delta}{\sin\delta_R^{EL}\cos\delta_B^{EL}}\,.
\label{eqn:dmwnoelle}
\end{equation}
Due to the explicit appearance of the background contributions, there is again 
no simple interpretation for $N(\mbox{Noelle})$ in terms of the dynamically
calculated quantities. On the other hand, the ratio $R_{EM}$ 
becomes quite simple
\begin{equation}
  R^R_{EM}(\mbox{Noelle}) = \frac{\bar
 t_{\pi\gamma}(E2)\cos\delta_B^{EL} - \bar t_{\pi\gamma}^B(E2)\cos
 \delta }{\bar t_{\pi\gamma}(M1)\cos\delta_B^{EL} - \bar
 t_{\pi\gamma}^B(M1)\cos \delta }\,.
\label{eqn:emrnoelle}
\end{equation}
Obviously, there is no simple relation between this ratio and
the dynamical resonance parameters. However,
if one evaluates this ratio at $\delta=\pi/2$, one finds 
\begin{equation}
  R^R_{EM}(\mbox{Noelle})|_{\delta=\pi/2} =
 R_{EM}(\mbox{DM})|_{\delta=\pi/2} \,.
\label{eqn:emrnoepi}
\end{equation}

In the K-matrix approach, one identifies the tree level contributions
as K-matrix elements. The full unitarized photoproduction amplitude reads 
then 
\begin{equation}
\bar t_{\pi\gamma} = (A_B + \frac{N}{\epsilon} )\cos\delta \,.
\end{equation}
Since here the following relation holds
\begin{equation}
\nonumber
\frac{1}{\epsilon} = \tan \delta -\tan \delta_B^{EL}\,,
\end{equation}
one finds
\begin{equation}
\bar t_{\pi\gamma}=A_B\cos \delta + \frac{N}{\cos \delta_B^{EL}}
  \sin \delta_R^{EL}\,.
\label{tpigkmatrix}
\end{equation}
Comparing (\ref{tpigkmatrix}) with (\ref{tpignoelle}) one finds 
the following relation 
\begin{equation}
N (\mbox{K-matrix}) = \cos \delta_B^{EL} N (\mbox{Noelle}),
\label{eqn:dmwkmat}
\end{equation}
which immediately leads to the same energy dependent 
$R^R_{EM}$ for the K-matrix as in 
Noelle's approach, and thus in the K-matrix approach the ratio quoted in 
the literature is nothing else than the ratio of the experimental multipoles
at $\delta = \pi/2$. 

We close this Section by comparing the explicit numerical values for
the various ratios $R_{EM}$ and other quantities introduced above with 
the dynamical model of Tanabe and Ohta (model B in \cite{TaO85}).
From their model one deduces from Eqs.\ (\ref{eqn:M_do})-(\ref{eqn:M_dk})
the following $\Delta$ masses: $M_\Delta \mbox{(Olsson)} = 1220\,
\mbox{MeV}$, $M_\Delta \mbox{(K-matrix)} = 1230\, \mbox{MeV}$ and 
$M_\Delta \mbox{(Noelle)} = 1243\, \mbox{MeV}$. The ordering of the
masses is due to the negative self-energy contribution of this
model. It turns out, that the values obtained in this manner
are not far away from the numbers quoted in \cite{DaM91}, which are
$1217\,$MeV, $1232\,$MeV, and $1250\,$MeV, respectively. The small
difference in the K-matrix method has its origin in the fact that
different data for $\delta$ had been fitted in both papers 
\cite{TaO85,DaM91}.

\begin{figure}
\centerline{\psfig{file=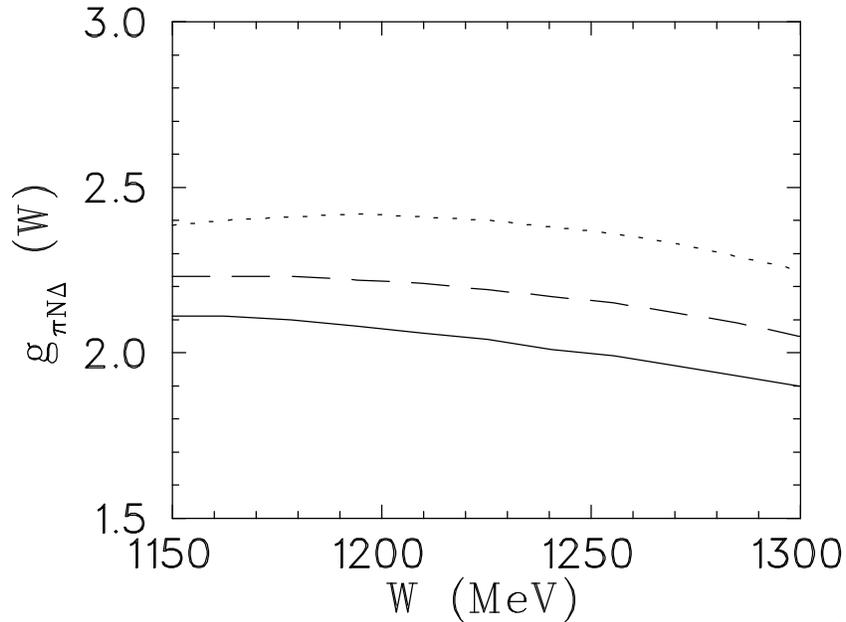,width=11cm,angle=0}}
\caption{The coupling strength $g_{\pi N\Delta}$ as function of the invariant
energy of the $\pi N$ system for the different ELs. The
dotted, dashed and solid curves correspond to the Olsson, Noelle and
K-matrix method, respectively. }
\label{fig:gpind}
\end{figure}

As next we calculate on the basis of the DM the coupling ``constant'' 
$g_{\pi N\Delta}(s)$ for the different ELs via the decay width 
$\Gamma_\Delta (s)$ given in (\ref{eqn:gamdef}) using Eqs.\ 
(\ref{eqn:epsdef}), (\ref{epsilons}) and (\ref{mdeltas}). The result is shown 
in Fig.\ \ref{fig:gpind}. One readily notices a 15 percent variation of the 
coupling strength between Olsson and K-matrix method. Furthermore, 
the energy dependence, which can be traced back to the energy dependence 
of the self-energy in the DM, is rather weak and similar for all three 
schemes but not negligible 
at all.  At least for the simple model of Tanabe and Ohta, it is
larger than the energy dependence allowed for by Davidson {\it et al.}
\cite{DaM91}.

\begin{figure}
\centerline{\psfig{file=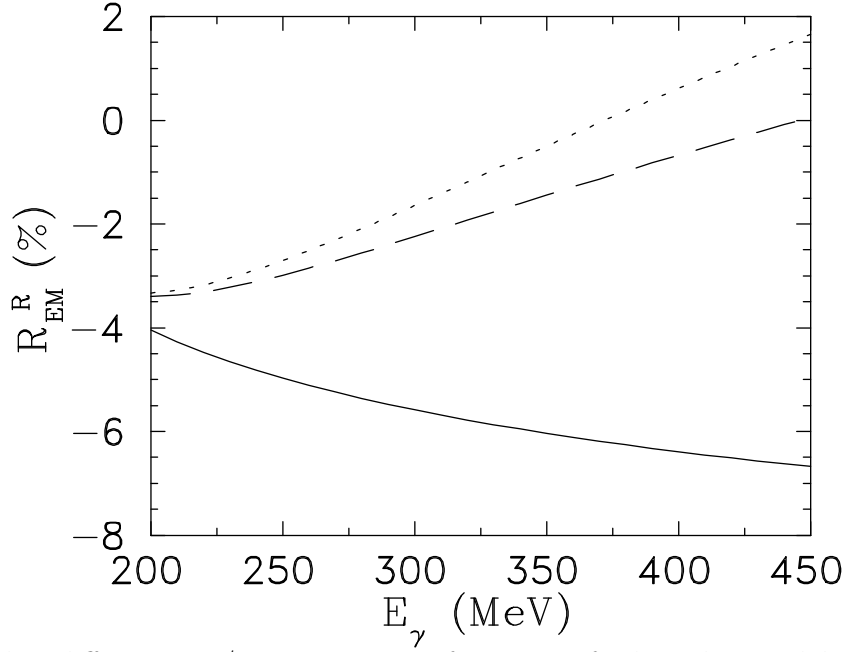,width=11cm,angle=0}}
\caption{The different E2/M1 ratios as
function of the photon laboratory energy: $\tilde R^R_{EM}\mbox{(DM)}=
R_{EM}^R$(Olsson) (solid), $\Re e [R_{EM}^R\mbox{(DM)}]$ (dashed), and 
$R_{EM}^R\mbox{(Noelle)} = R_{EM}^R$(K-matrix) (dotted).}
\label{fig:e2m1}
\end{figure}

In Fig.\ \ref{fig:e2m1} we show the various E2/M1 ratios as a function
of the photon lab energy.  There are basically three different dressed
ratios, $\tilde R^R_{EM}\mbox{(DM)}= R_{EM}^R$(Olsson),
$\Re e[R_{EM}^R(\mbox{DM})]$, 
and $R_{EM}^R \mbox{(Noelle)} = R_{EM}^R$(K-matrix).
Note that these identifications are based on the assumption that the
background contributions, which may formally be obtained by putting $g_{\pi
N\Delta}= 0 $ in DM and EL, can be identified.  The difference between
the ratios is obvious. In particular, comparing $R_{EM}^R$(Olsson) with 
$\Re e[R_{EM}^R(\mbox{DM})]$, 
one readily notes quite a distinct different energy 
dependence, while $R_{EM}^R$(K-matrix) behaves qualitatively similar 
as $\Re e[R_{EM}^R(\mbox{DM})]$, although differences remain in 
detail. If evaluated at the respective masses
$M_\Delta$, we find $R^R_{EM}\mbox{(Olsson)} = -5.7$\%,
$R^R_{EM}\mbox{(Noelle)} = -0.2$\% and $R^R_{EM}\mbox{(K-matrix)} =
-0.7$\%. The difference to the results of \cite{DaM91} can be traced 
back mainly to different prescriptions for the e.m.\ and hadronic 
background contributions. Since all ratios are based on the same DM, these 
different results clearly reflect the fact that a meaningful comparison is 
not possible. In this context we recall that except for $R_{EM}^R$(K-matrix), 
the dressed ratios are representation dependent.

\section{Speed plot analysis}
\label{sect4}

The essential idea of the speed plot analysis (SP) rests on the
analytic properties of the partial wave amplitude. 
A resonance manifests itself as a pole in the
lower half plane of the second Riemann sheet when continuing the
amplitude into the complex energy plane.  The aim of the SP is to
locate the pole position and to determine the corresponding residue
from the experimental data.  This method, originally used in pion
scattering, has been applied very recently to pion photoproduction 
by Hanstein, Drechsel and
Tiator \cite{HaD96}. Assuming a well isolated pole for the $\Delta$
resonance, one can separate it by performing a Laurent
expansion in the vicinity of the pole, resulting in the form
\begin{equation}
t_{\pi\gamma}(W)=t_{pole}^{SP}(W)+t_{regular}^{SP}(W)\,,\label{spamp}
\end{equation}
with the pole term
\begin{equation}
t_{pole}^{SP}(W)= \frac{r\,e^{i\phi}\Gamma_R /2}{W_R -W}\,,
\end{equation}
and the regular part
\begin{equation}
t_{regular}^{SP}(W)=\sum_{n=0}^{\infty} \tau_n (W-W_R)^n\,,
\end{equation}
with constant complex coefficients $\tau_n$.
Here, $W_R=M_R-\frac{i}{2}\Gamma_R$ denotes the pole position with constant 
resonance mass $M_R$ and width $\Gamma_R$, and $r$ and $\phi$ characterize 
modulus and phase of the residue. 

If we apply this analysis to the DM, the pole is determined as solution of
\begin{equation}
z=m_\Delta^0+\Sigma_\Delta(z)
\end{equation}
on the second Riemann sheet, which leads to a sytem of coupled 
equations for the resonance mass and width
\begin{eqnarray}
M_R&=&m_\Delta^0+\Re e\,[\Sigma_\Delta(M_R-\frac{i}{2}\Gamma_R)]\,,\\
\Gamma_R &=& -2\,\Im m\,[\Sigma_\Delta(M_R-\frac{i}{2}\Gamma_R)]\,.
\end{eqnarray}
Expanding then the DM amplitude as in (\ref{spamp}), one immediately
sees that the splitting into a pole and a regular part formally is not
identical with the splitting into a resonance and a background part
because of an additional energy dependence of the dressed resonance
amplitude besides the pole contribution.

However, this formal criticism may not be too relevant in the actual 
application if the background amplitude is very weakly energy dependent. 
To this end, 
let us first briefly review the SP of elastic
$\pi N$ scattering as performed by H\"ohler \cite{HoS92,Hoe93} to determine
the resonance pole parameters ($M_R$, $\Gamma_R$). It is based on the
following parametrization of a resonant partial wave amplitude $t_{\pi\pi}$
by splitting it into a background $t_{\pi\pi}^b$ and a pole 
part $t_{\pi\pi}^{pole}$ 
\begin{equation}
  \label{hoehler_ansatz}
  t_{\pi\pi}(W) = t_{\pi\pi}^b(W) + t_{\pi\pi}^{pole}(W)\,,
\end{equation}
where
\begin{equation}
t_{\pi\pi}^{pole}(W) = z(W)t_{\pi\pi}^r(W)\,,
\end{equation}
with
\begin{equation}
  t_{\pi\pi}^r(W)=\frac{\Gamma_R/2}{M_R-W-i\Gamma_R/2}\,.
\end{equation}
Since the $\Delta$ resonance lies in the elastic region, H\"ohler's
parametrization requires in addition separate unitarity for the background. 
Thus a real phase shift $\delta_b$ can be assigned by
\begin{equation}
  \label{tb}
  t^b_{\pi\pi}(W) = \sin\delta_b(W) e^{i\delta_b(W)}.
\end{equation}
As discussed in Sect.\ II, Eq.\ (\ref{tb}) is a natural consequence in
any dynamical model. 
In the elastic region, of course, also the full amplitude is unitary,
\begin{equation}
  \label{tfull}
  t_{\pi\pi}(W) = \sin\delta(W) e^{i\delta(W)},
\end{equation}
which fixes the quantity $z$ 
\begin{equation}
z(W)=e^{2i\delta_b(W)}.  
\end{equation}

Assuming $t_{\pi\pi}^b$ (and thus $\delta_b$ and $z$) to be energy
independent, one can extract the pole parameters $M_R$ and $\Gamma_R$
directly from the speed of the full amplitude, which is defined as
\begin{equation}
  \label{eq:speed}
  SP_{\pi\pi}(W)=\left|\frac{dt_{\pi\pi}(W)}{dW}\right|.
\end{equation}
Then one finds from (\ref{hoehler_ansatz})
\begin{equation}
SP_{\pi\pi}(W)= \frac{z\Gamma_R/2}{(M_R-W)^2+\Gamma_R^2/4}\,.
\end{equation}
Thus the maximum of the speed is located at $W=M_R$, determining the
resonance mass. Furthermore, the width is then obtained from the half maximum
values
\begin{equation}
SP_{\pi\pi}(M_R\pm\Gamma_R/2)=SP_{\pi\pi}(M_R)/2\,,
\end{equation}
or from the curvature at maximum speed
\begin{equation}
\Gamma_R^2
=-8 \frac{ SP_{\pi\pi}(M_R)} {d^2SP_{\pi\pi}(M_R)/dW^2} \,.
\end{equation}
The latter should be preferred for a broad resonance but might be more
difficult to evaluate numerically.  H\"ohler \cite{HoS92} finds
\begin{equation}
  \label{hoehler_result}
  M_R=1210\,\mbox{MeV}, \quad\quad\Gamma_R=100\,\mbox{MeV}.
\end{equation}
These values may be used to calculate $\delta_b$ from
(\ref{hoehler_ansatz}) for the whole energy range and to check the
initial assumption with respect to the energy dependence of
$t_{\pi\pi}^b$.  Indeed, one finds at resonance
$\delta_b(W=M_R)=-23.5^\circ$ and only a very small variation of
$\delta_b$ with energy which justifies the method. 
An important point to note so far is that $t_{\pi\pi}^b$ and
$t_{\pi\pi}^{pole}$ in H\"ohler's analysis may well be considered as
approximations to background and dressed amplitudes $t_{\pi\pi}^B$ and
$t_{\pi\pi}^R$ of a certain dynamical model in a limited energy
intervall around the resonance position.  In view of the fact that one
can generate a whole family of phase equivalent dynamical models by
means of unitary transformations without changing the pole position in
the complex plane, one may conclude that the speed plot analysis
singles out from this family the one which predicts a dressed
amplitude $t_{\pi\pi}^R$ that is most accurately approximated by its
pole term and which gives the weakest energy dependence of the
background $t_{\pi\pi}^B$. On the mass shell, it would provide dynamical 
amplitudes $t_{\pi\pi}^B$ and $t_{\pi\pi}^R$ which are closest to H\"ohler's
amplitudes $t_{\pi\pi}^b$ and $t_{\pi\pi}^{pole}$, respectively.
This observation could also provide a possible explanation for the fact 
that the resulting background phase is repulsive in contrast to standard 
attractive background interactions as has been noted already by H\"ohler 
\cite{Hoe93}.

The recent extension of the SP to pion photoproduction by Hanstein {\em et 
al.} \cite{HaD96} takes as starting point the following ansatz for the 
photoproduction amplitude 
\begin{equation}
  \label{hanstein_ansatz}
  t_{\pi\gamma}(W) = t_{\pi\gamma}^{born}(W) 
+ \tilde t^{\,b}_{\pi\gamma} + z_\gamma t_{\pi\pi}^r(W)\,,
\end{equation}
where $t_{\pi\gamma}^{born}$ is the photoproduction Born amplitude
calculated from the pseudoscalar $\pi N$ Lagrangian.  Assuming $\tilde
t^{\,b}_{\pi\gamma}$ to be energy independent, they could determine the
pole parameters $M_R$, $\Gamma_R$ and the residue $z_\gamma$ from the
speed of the difference between the full and the Born amplitude,
\begin{equation}
  \label{speed_ha}
SP_{\pi \gamma}(\mbox{Hanstein})= 
 \left| \frac{d\,[t_{\pi\gamma}(W)-t_{\pi\gamma}^{born}(W)]}{dW} \right| .    
\end{equation}

The main reason for critisizing this procedure is two-fold: ({\em i}) In
the original SP of $\pi N$ scattering
\cite{HoS92,Hoe93} Born terms have not been subtracted, i.e., the
speed of the full amplitude determines the pole parameters. Moreover, only the
speed of the full amplitude is related to the time delay between the
arrival of the incident wave packet at the collision region and the
departure of the outgoing packet. ({\em ii}) There is no reason to
give up the separate unitarity constraint for the background amplitude
and to treat the hadronic and e.m.\ reactions in a
different manner. 

Therefore, we propose as a consequent extension of H\"ohler's ansatz 
(\ref{hoehler_ansatz}) to e.m.\ pion production in the elastic region 
the form 
\begin{equation}
  \label{u_ansatz}
  t_{\pi\gamma}(W) = \beta(W) e^{i\delta_b(W)} + z_\gamma(W) t_{\pi\pi}^r(W),
\end{equation}
where again it is assumed that the background term $\beta 
e^{i\delta_b}$ obeys itself the unitarity constraint, i.e., it
fulfills Watson's theorem which implies that $\beta$ is real.
The constraint from Watson's theorem for the full amplitude
$t_{\pi\gamma}$, which can be written as
\begin{equation}
  \tan\delta(W) = \frac {\Im m (t_{\pi\gamma}(W))} {\Re e
    (t_{\pi\gamma}(W))},
\end{equation}
finally allows to eliminate the unknown modulus of the background
amplitude $\beta(W)$ in Eq.\ (\ref{u_ansatz}). One finds
\begin{equation}
  \label{speed_watson}
  \beta(W) = -\sin\delta_b(W)\Re e\,[z_\gamma(W)]
             +\cos\delta_b(W)\Im m\,[z_\gamma(W)].
\end{equation}
Of course, the
information contained in (\ref{u_ansatz}) and (\ref{speed_watson}) is
not sufficient to fix the complex function $z_\gamma(W)$ over a larger
energy range from available experimental multipole data sets.
However, it allows to check whether the data can be described at all
over a limited energy range around the resonance position replacing
$z_\gamma(W)$ by the residue $z_\gamma(M_R)$. 

\begin{figure}
\centerline{\psfig{file=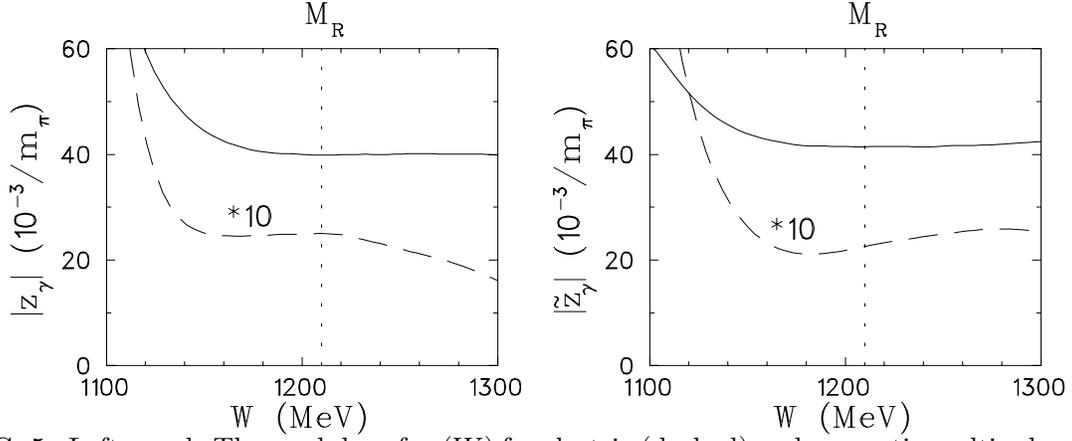,width=14cm,angle=0}}
\caption{Left panel: The modulus of \protect{$z_\gamma(W) $} for 
electric (dashed) and magnetic multipoles (solid). 
Right panel: The modulus of $\tilde z_\gamma$ of mSP.}
\label{fig:residue}
\end{figure}

We have performed such a SP taking as input ``data'' the multipoles from
the fixed-$t$ dispersion analysis of Hanstein et al.\ \cite{HaD96}. By
construction, these multipoles exactly obey Watson's theorem, and thus
they provide at the same time a ``data'' set for 
the $ \pi N$ scattering phase shift
$\delta$. 
Writing the residue $z_\gamma(M_R)=r\exp(i\phi)$, it is obvious (see 
(\ref{u_ansatz}) and (\ref{speed_watson})) that its modulus $r$ is given 
by the ratio of the speeds of the hadronic and
e.m.\ amplitudes
\begin{equation}
  \label{eq:resr}
  r= \frac{|dt_{\pi\gamma}/dW|}{|dt_{\pi\pi}/dW|}\,. 
\end{equation}
This ratio is plotted as a function of the energy in the left panel 
of Fig.\ \ref{fig:residue}. The plateau around $W=M_R$ already indicates 
that the assumption of a constant residue may provide a good approximation 
in an energy interval around the resonance position. The phase $\phi$ of the 
residue can be calculated from the relation 
\begin{equation}
  \label{eq:resphi}
  \Im m\,[ t_{\pi\gamma}(M_R)] = \frac{r}{1+\tan^2\delta_b}
\Big[ \cos\phi +\tan\delta_b\sin\phi \Big]\,,
\end{equation}
which follows directly from (\ref{u_ansatz}) and (\ref{speed_watson}). 
Our values for $z_\gamma(M_R)$ are compared with the ones of Ref.\ 
\cite{HaD96} in Table \ref{tab1}.  In the case of the
electric multipole $E_{1+}^{3/2}$, both values agree very well.
However, a somewhat larger deviation is found for $M_{1+}^{3/2}$.  For the
ratio of the residues of electric to magnetic multipoles we obtain
\begin{equation}
  \label{eq:zgamma}
 R_{EM}(\mbox{SP})=  \frac{z^E_\gamma(M_R)}{z^M_\gamma(M_R)} 
                  = -0.040-0.047\,i\,.
\end{equation}
Comparison to the result of \cite{HaD96} $(-0.035-0.046i)$ shows a good 
agreement for the imaginary part while the real part is 10 percent smaller. 

\begin{figure}
\centerline{\psfig{file=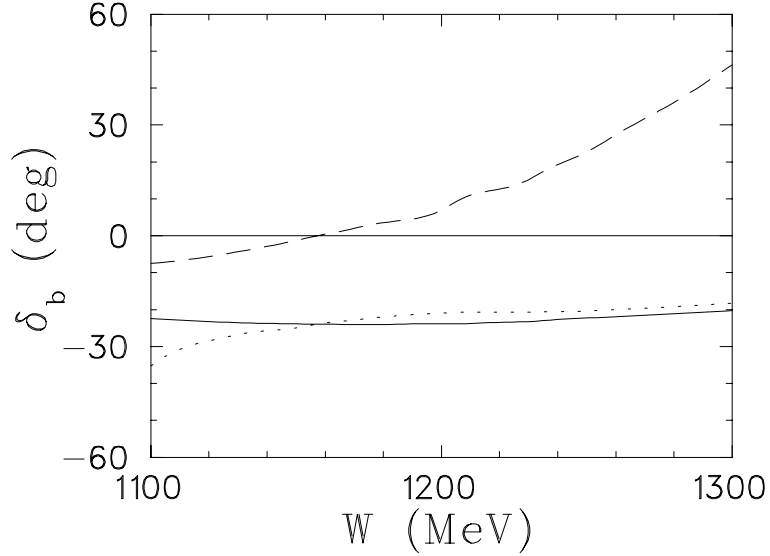,width=10cm,angle=0}}
\caption{Background phase shifts as function of the photon lab energy 
  from the analysis of \protect\cite{HaD96}. The dotted and
  dashed curves correspond to the electric and magnetic multipoles,
  respectively. The solid curve shows the background phase obtained
  by H\"ohler from \protect{$\pi N$} scattering \protect\cite{HoS92}.}
\label{fig:speedphases}
\end{figure}

\begin{figure}
\centerline{\psfig{file=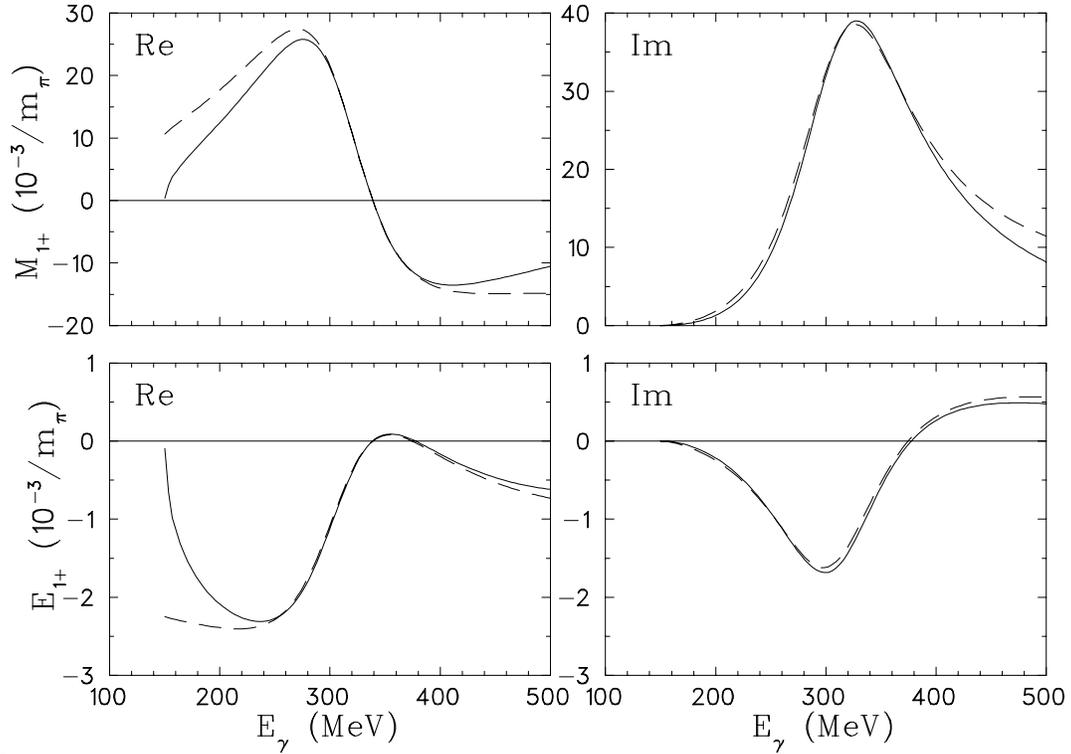,width=14cm,angle=90}}
\caption{The pole approximation as obtained in our analysis (dashed) 
  compared to the complete multipoles from Hanstein {\it et al.}
  \protect\cite{HaD96} (solid).}
\label{fig:polapprox}
\end{figure}

Most of the difference between the approach of Hanstein {\em et al.}
and ours can be traced back to the phase of the background of the
magnetic multipole.  Whereas in our ansatz the phases of both
background amplitudes agree by definition with $\delta_b$, which is
already fixed in the analysis of $\pi N$ scattering, the phase of the
background $t_{\pi\gamma}^{born} + \tilde t^{b}_{\pi\gamma}$ in
(\ref{hanstein_ansatz}) is unconstrained since $\tilde
t^{b}_{\pi\gamma}$ is an arbitrary complex number. As shown in Fig.\
\ref{fig:speedphases}, they differ even in sign for both multipoles.
Only the electric background phase (dotted curve) is close to
$\delta_b$ (solid curve).  Consequently, the background and pole terms
found in \cite{HaD96} cannot, even approximately, be identified with
the background and dressed multipoles $t^B_{\pi\gamma}$ and
$t^R_{\pi\gamma}$, respectively, of any dynamical model.  However,
this is possible for our results in the approximate way discussed
above.  Fig.\ \ref{fig:polapprox} shows the quality of the pole
approximation, i.e., evaluating (\ref{u_ansatz}) with
$z_\gamma(W)=z_\gamma(M_R)$, in comparison with the full
multipoles. Clearly, the pole approximation cannot reproduce the data
below and above the resonance region as it not need to.  In
particular, it is obvious that by definition the pole approximation
leads to a wrong low-energy behaviour.

There remains one final point to be mentioned.  It concerns the proper
normalization of the amplitude for which the speed is calculated since
any energy dependent normalization factor would affect the result.
According to \cite{Hoe93}, two times the speed of the $T$-matrix (which is
related to the $S$-matrix through $S=1+2iT$) defines the quantity
which can be interpreted as a time delay.  This suggests to study the
speed of the product of $\sqrt{qk}$ times the multipole, where $q$ and
$k$ are the c.m.\ momenta of pion and photon, respectively. (The
relation between $T$-matrix and multipoles can be found e.g.\ in
\cite{BeM92}). Since $d\sqrt{qk}/dW\approx 0.9$ in the resonance region, 
this factor is not negligible.  We have performed this modified speed
analysis (mSP), i.e., identified $t_{\pi\gamma}$ in Eq.\ 
(\ref{u_ansatz}) with $\sqrt{qk}$ times the multipole, instead of taking 
the multipole itself as before and in Ref.\ \cite{HaD96}.  The
results are summarized in Fig.\ \ref{fig:residue} (right panel) and
Table \ref{tab1} (middle column).  In order to allow a direct
comparison with the foregoing SP, we present the mSP results in terms of 
$\tilde z_\gamma\equiv z_\gamma/\sqrt{q_0k_0}$ where $q_0$ and $k_0$ are 
the momenta at $W=M_R$.  From Table \ref{tab1} one obtains
\begin{equation}
  R_{EM}(\mbox{mSP}) = -0.032 -  0.044\,i\,,
\end{equation}
which differs significantly from $R_{EM}(SP)$ in (\ref{eq:zgamma}).  
In addition, we have checked that the
quality of the pole approximation, which can be achieved within the mSP, 
is very similar to the one shown in Fig.\ \ref{fig:polapprox}.

This last point shows the following: Without reference to the
physical interpretation of the speed in terms of the above mentioned 
time delay, one would not have an unique prescription for the choice 
of the amplitude for which the speed should be calculated and thus 
the whole procedure would become rather ambiguous. 

\section{Summary and conclusions}\label{sect5}

We have analyzed in detail different theoretical descriptions of pion 
photoproduction in the region of the $\Delta$ resonance, namely, dynamical 
models, effective Lagrangian approaches, and the recently proposed speed 
plot analysis. Our main emphasis has been laid on the question to what 
extent genuine resonance contributions to the e.m.\ multipoles can be 
extracted, in particular, the interesting and controversely discussed 
E2/M1 ratio $R_{EM}$ for the e.m.\ $\Delta$ excitation. 

As theoretical basis we have chosen a dynamical model for which we have 
briefly reviewed the treatment of pion nucleon scattering and pion 
photoproduction, in particular the separation of background and resonance 
properties. For the Effective Lagrangian approaches, we have discussed the 
different unitarization schemes resulting in different background and 
resonance contributions as pointed out already in \cite{DaM91}. 
It turns out that a priori there is no direct relation of background 
and resonance contributions of a DM to the corresponding ones of a EL. 
The main reason for this is that unitarity is put in by hand in the EL 
which does not allow the identification of the background part. Only on the 
premises that the hadronic background phases could be identified, 
the various resonance 
parameters like resonance mass, decay width, and E2/M1 ratio could be 
related. This reflects the fact, that the largely phenomenological 
treatment of the background introduces a unitary ambiguity or 
representation dependence \cite{WiW96}. Therefore, almost all E2/M1 
ratios discussed so far are representation dependent quantities 
which implies that they are not observable in the strict sense. 
There is one exception, namely the ratios in
the Noelle and K-matrix approach, when evaluated at the resonance
position, because they are 
simply given by the ratio of the full multipoles at this energy. 

The main problem, however, how these numbers should be compared to a 
given microscopic hadron model, remains unresolved.
The only safe solution is to calculate the complete pion production 
amplitude within such a model in order to have an unambiguous 
and direct comparison to experimentally observable quantities. 

As last topic, we have studied the recently proposed speed
plot analysis of Hanstein {\it el al.} We could show that the
separated resonance and background contributions of their analysis
cannot be identified with corresponding DM quantities, since their ansatz 
for the background is not constrained by unitarity. For this reason, 
we have proposed an alternative speed plot analysis
which respects the separate unitarity condition for the background. 
In such an analysis, 
the regular and pole contributions can be viewed as
approximations of the background and dressed resonant contribution of
a certain dynamical model.  
Applied to the same input as Hanstein {\it et al.}, we find 
a similar result for  the residue of the electric multipole, whereas 
differences of the order of 10 percent appear for the magnetic multipole. 
Accordingly, we obtain a slightly changed ratio 
$R_{EM}(\mbox{SP})$. In addition, we have proposed a further modification 
of the speed analysis which was guided by the 
physical interpretation of the speed in terms of the time delay between 
incoming and outgoing wavepackets. This modification leads to a 20 
percent change in the $R_{EM}(\mbox{SP})$ ratio. 
But again it remains unclear how 
$R_{EM}(\mbox{SP})$ could be compared to any microscopic hadron model. 

Even though we have focused here on the
$\gamma N\leftrightarrow \Delta$, the conclusions are also valid for
the case of electroexcitation and/or the excitation of higher
resonances. In the latter case, the situation is even worse due to 
overlapping resonances and the
coupling to many pion channels, which makes a realistic dynamical
calculation more difficult.

\section*{Acknowledgements}

We thank R.\ Beck and A.\ M.\ Bernstein for many fruitful discussions.
This work was supported by the Deutsche Forschungsgemeinschaft (SFB
201).

\begin{table}
\caption{
  Residues $z_\gamma(M_R)=r\exp(i\phi)$ for the $E_{1+}^{3/2}$ and
 $M_{1+}^{3/2}$ multipoles from various speed plots.}
\label{tab1}
\begin{tabular}{lccc}
  & present (SP) & present (mSP) & from \protect\cite{HaD96} \\ 
\hline
  $r_E$ $(10^{-3}/m_\pi)$ & 2.47 & 2.24 & 2.46 \\
  $\phi_E$ (deg) & $-$156 & $-162$ & $-$154.7 \\
  $r_M$ $(10^{-3}/m_\pi)$ & 39.9 & 41.5 & 42.32 \\
  $\phi_M$ (deg) & $-$26.0 & $-36.5$ & $-$27.5 \\
\end{tabular}
\end{table}

\end{document}